\documentclass[aps,pra,twocolumn,superscriptaddress,showpacs,amsmath,amssymb,10pt]{revtex4-1}
\usepackage[english]{babel}
\usepackage{amsmath}
\usepackage{bm}
\usepackage{graphicx}
\usepackage{bbm} 
\usepackage{times}
\usepackage{epsfig} 
\usepackage[colorlinks,linkcolor=blue,citecolor=blue,urlcolor=black]{hyperref}
\begin{document}

\title{Efficiency of fermionic quantum distillation}

\author{J. Herbrych}
\affiliation{Department of Physics and Astronomy, University of Tennessee, Knoxville, Tennessee 37996, USA}
\affiliation{Materials Science and Technology Division, Oak Ridge National Laboratory, Oak Ridge, Tennessee 37831, USA}
\author{A. E. Feiguin}
\affiliation{Department of Physics, Northeastern University, Boston, Massachusetts 02115, USA}
\author{E. Dagotto}
\affiliation{Department of Physics and Astronomy, University of Tennessee, Knoxville, Tennessee 37996, USA}
\affiliation{Materials Science and Technology Division, Oak Ridge National Laboratory, Oak Ridge, Tennessee 37831, USA}
\author{F. Heidrich-Meisner}
\affiliation{Department of Physics and Arnold Sommerfeld Center for Theoretical Physics, Ludwig-Maximilians-Universit\"{a}t M\"{u}nchen, 80333 M\"{u}nchen, Germany}

\date{\today}

\begin{abstract}
We present a time-dependent density-matrix renormalization group investigation of the quantum distillation process within the Fermi--Hubbard model on a quasi-1D ladder geometry. The term distillation refers to the dynamical, spatial separation of singlons and doublons in the sudden expansion of interacting particles in an optical lattice, i.e., the release of a cloud of atoms from a trapping potential. Remarkably, quantum distillation can lead to a contraction of the doublon cloud, resulting in an increased density of the doublons in the core region compared to the initial state. As a main result, we show that this phenomenon is not limited to chains that were previously studied. Interestingly, there are additional dynamical processes on the two-leg ladder such as density oscillations and selftrapping of defects that lead to a less efficient distillation process. An investigation of the time evolution starting from product states provides an explanation for this behaviour. Initial product states are also considered, since in optical lattice experiments such states are often used as the initial setup. We propose configurations that lead to a fast and efficient quantum distillation.
\end{abstract}
\maketitle

\section{Introduction}
\label{sec:intro}

The interest in the nonequilibrium dynamics of interacting quantum many-body systems has been driven both by recent experiments and by theoretical considerations \cite{Polkovnikov2011,Eisert2015,Langen2015,Gogolin2015}. On the experimental side, we highlight the possibility to study the quantum-quench dynamics of ultra-cold atomic gases in optical lattices \cite{Greiner2002a,Trotzky2012}. While many seminal experiments focused on Bose gases \cite{Kinoshita2006,Hofferberth2007,Gring2012,Cheneau2012,Langen2013,Kaufman2016}, the successful implementation of fermionic quantum-gas microscopes by a large number of experimental groups will likely draw future attention to fermions \cite{Cheuk2015,edge2015,Omran2015,Parsons2015,haller2015,greif2016,Boll2016,Cocchi2016,Cheuk2016a,Cheuk2016,Parsons2016,Mitra2017}, further adding to the existing experimental work on quench dynamics in fermionic lattice gases \cite{Schneider2012,Pertot2014,Will2015,Schreiber2015}. In parallel, the marriage of pump-and-probe spectroscopy and strongly-correlated electron systems is rendering the investigation of ultrafast dynamics of correlated electrons a timely topic in condensed matter physics as well \cite{Orenstein2012,Giannetti2016,Gandolfi2016}. 

In the field of ultra-cold quantum gases, a significant amount of experimental work concentrates on the relaxation and thermalization dynamics of low-dimensional quantum systems \cite{Kinoshita2006,Hofferberth2007,Cheneau2012,Gring2012,Langen2013,Will2015,Kaufman2016,Schreiber2015,Choi2016}. In parallel, experiments focusing on the nonequilibrium transport properties of atoms in optical lattices were pushed forward, ranging from the few-body \cite{Fukuhara2013,Fukuhara2013a,Preiss2015} to the many-body regime \cite{Schneider2012,Ronzheimer2013,Xia2014,Vidmar2015}. Not surprisingly, many unusual and sometimes counterintuitive phenomena exist in the {\it transient} dynamics of nonequilibrium problems, such as prethermalization \cite{Gring2012,Berges2004,Moeckel2008,Eckstein2009}, the dynamical quasi-condensation of hard-core bosons \cite{Rigol2004,Rigol2005a,Rodriguez2006,Vidmar2013,Vidmar2015}, or the quantum distillation mechanism \cite{Heidrich-Meisner2009,Bolech2012,Xia2014}.

\begin{figure}[!t]
\includegraphics[width=1.0\columnwidth]{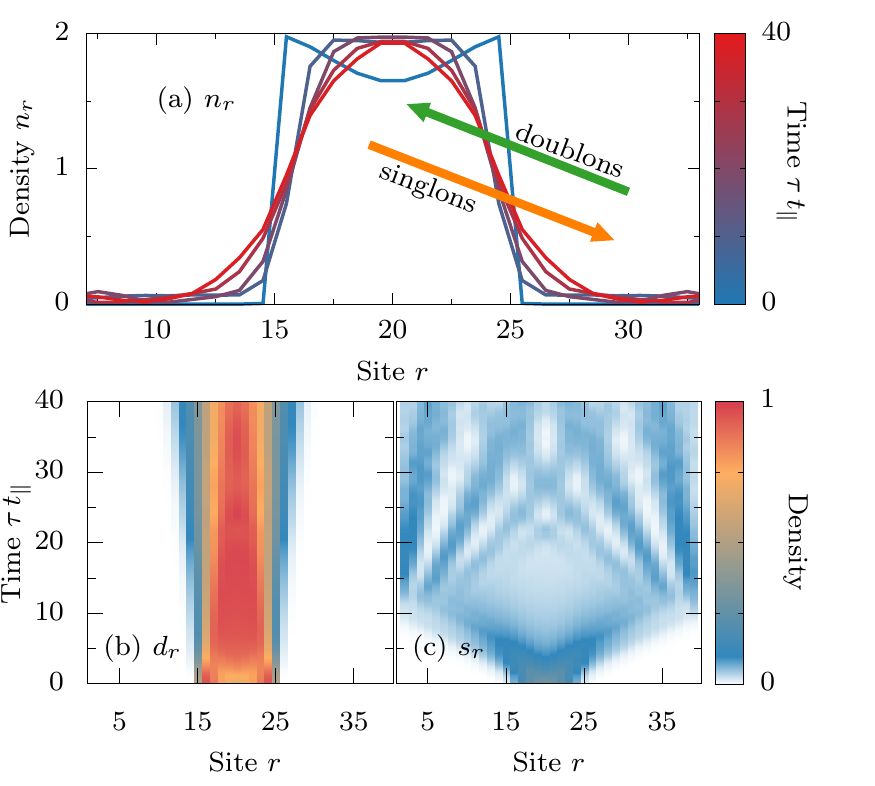}
\caption{(Color online) (a) Snapshots of the time evolution of the local density profiles $n_{r}$ as calculated for a single chain with $U/t_{\parallel}=40$, $n_{\mathrm{conf}}=1.8$, and $N=18$. (b,c) Time evolution of (b) doublon $d_{r}$ and (c) singlon $s_{r}$ densities for the same parameters as in panel (a). See the text for details.}
\label{fig1}
\end{figure}

Our work will focus on such an aspect of transient nonequilibrium mass transport, namely the quantum distillation mechanism in a system of interacting fermions on an optical lattice. Quantum distillation is the dynamical spatial separation of the lattice gas into one portion that carries predominantly interaction energy and another one that carries mostly kinetic energy. This spatial separation occurs during the so-called sudden expansion \cite{Schneider2012,Ronzheimer2013,Xia2014,Vidmar2015}, i.e., the release of an interacting quantum gas from a trap and its subsequent expansion into an empty optical lattice (see \cite{Rigol2004,Rigol2005,Rigol2005a,Minguzzi2005,Rodriguez2006,Heidrich-Meisner2008,Heidrich-Meisner2009,Hen2010,Kajala2011,Langer2012,Bolech2012,Jreissaty2011,Jreissaty2013,Vidmar2013,Boschi2014,Campbell2015,Hauschild2015,Schluenzen2016,Mei2016,Vidmar2017,Vidmar2017a,Xu2017} for theory work on this specific nonequilibrium problem). In the presence of strong on-site interactions $U$ much larger than the typical tunneling matrix element $t$, particles of opposite spin bound into a doublon on the same site can only move with an effective tunneling matrix element $t_{d} \sim 4t^2/|U|$. Moreover, these doublons are dynamically stable over an exponentially long time \cite{Winkler2006,Rosch2008}. Therefore, in an initial state that has a large contribution of doublons, the cloud can only expand on time scales proportional to $1/t$ and via first-order processes if doublons exchange their position with neighboring singlons, resulting in the doublons moving towards the core of the system and allowing the singlons to expand \cite{Heidrich-Meisner2009}. As a consequence of this transient dynamics, the core region of the system will mainly contain doublons and hence $n_r \approx 2d_r$ (where $n_r$ and $ d_r$ are the particle and doublon density at given site $r$, respectively) while in the expanding wings, $n_r \approx s_r$ with $s_r$ the singlon density, and thus a spatial separation of regions of high interaction versus high kinetic energy occurs. The effect crucially relies on energy conservation and the bounded energy spectrum for a single-band system. A typical situation is illustrated in Fig.~\ref{fig1} for a Hubbard chain. Figure~\ref{fig1}(a) shows local densities at several times, with the singlons evaporating and the doublon density increasing in the core region. Figures~\ref{fig1}(b) and (c) show $d_r$ and $s_r$ as a function of time and position, confirming this picture. Note that a dynamical freezing of mass transport due to large density gradients in interacting lattice gases, also known as selftrapping (see, e.g., \cite{Anker2005,Reinhard2013}), already exists in the mean-field regime (see, e.g., \cite{Jreissaty2013}). Quantum distillation, however, goes beyond mere selftrapping and predicts an expansion of the initial cloud and the possible contraction of the doublon-cloud radius \cite{Heidrich-Meisner2009}.

The quantum distillation mechanism has first been proposed for interacting fermions in one-dimensional optical lattices \cite{Heidrich-Meisner2009} but works for bosons as well \cite{Muth2012}. We wish to distinguish between a strong and a weak version of quantum distillation: in the former, the density of doublons grows in the center of the system and even  the total local density increases beyond its initial value. In this regime, even a core region with a perfect band insulator can be produced with fermions, which clearly is the most extreme case of a very low-entropy region spatially separated from the high-entropy expanding singlons \cite{Heidrich-Meisner2009}. The strong version of quantum distillation has been suggested as a possible cooling mechanism for fermions \cite{Heidrich-Meisner2009}. The latter is an important goal in future fermionic optical lattice experiments \cite{Mckay2011} (see \cite{Joerdens2008,Schneider2008,Greif2014,Hart2015,Cheuk2016,Boll2016,Mazurenk2016} for recent advances).

The weak version of quantum distillation consists of a mere dynamical separation of singlons from doublons without an increase of the core particle density beyond its initial value. The separation between the two regimes, weak and strong quantum distillation, is a smooth crossover that depends on the initial conditions such as interaction strength or the shape of the confining potential \cite{Heidrich-Meisner2009}. A particular clean separation of a quantum gas into a {\it noninteracting} gas of singlons and a core region of pairs (which carry the interaction energy) can be accomplished by starting from a partially polarized gas with attractive interactions \cite{Bolech2012}. In that case, all minority fermions (say of spin $\sigma=\downarrow$) are bound into pairs in the strongly interacting regime and hence the singlons that evaporate out of the initially confined gas carry only $\sigma =\uparrow$ and form a non-interacting gas.

The quantum distillation effect was experimentally observed using bosons in a one-dimensional lattice \cite{Xia2014} and the separation of the expanding cloud into regions with predominantly singlons or doublons (and local objects formed of more than two bosons) was beautifully demonstrated. This experiment operated in the weak-quantum distillation regime with no discernible increase beyond the initial density. Moreover, for bosons, even a state of two particles per site would, in most cases, still be a correlated Mott insulator, while a fermionic state with two fermions (of opposite spin projection) per site is necessarily a product state. Thus, the experimental observation of quantum distillation with fermions and accessing the strong quantum distillation regime remain open. In a broader sense, the quantum distillation is one out of many interesting phenomena related to the presence of long-lived and heavy multi-particle objects in a {\it sea of singlons}, studied experimentally \cite{Winkler2006,Meinert2013,Xia2014} and theoretically \cite{Kollath2007,Petrosyan2007,Rosch2008,Pinto2009,Kessler2013,Rausch2017,Rausch2016}.

A goal of our work is to improve the understanding of fermionic quantum distillation and its efficiency. Before summarizing our main results, let us describe what we mean by the {\it efficiency} of the process. The perfect strong distillation leads to a band insulator in the core. The size of this dynamically formed band insulator is controlled by the initial number of doublons in the confined region \cite{Heidrich-Meisner2009}. The next criterion for the  efficiency is the largest value of the local doublon occupancy (or the particle density) reached in the expansion in the center of the system. Another important aspect of the quantum distillation is the time scale $\tau_{\rm QD}$ on which that maximum value is reached. Finally, even if strong quantum distillation is not realized, the desired feature is a fast dynamical spatial separation of singlons and doublons.

We study these aspects by pursuing three directions. First, and most importantly, the effect has almost exclusively  been discussed for one-dimensional systems (with the exception of a time-dependent Gutzwiller ansatz study of bosons in two dimensions \cite{Jreissaty2013}), where one can expect quantum distillation to be the most efficient: A singlon can only pass a neighboring doublon by exchanging the position with that doublon, while in a two-dimensional lattice, one could imagine percolation effects with singlons escaping via random paths of neighboring singlons and holons (empty sites). In order to address the question of the efficiency of quantum distillation beyond strictly one dimension, we consider the Fermi-Hubbard model on a {\it two-leg ladder} [see Fig.~\ref{fig2}(a)] and study the expansion dynamics as a function of the ratio $t_\perp/t_\parallel$, where $t_\parallel$ and $t_\perp$ denote the tunneling-matrix elements along the legs and rungs, respectively. We observe that the strong version of quantum distillation can be found on such ladders in a wide parameter regime, i.e., as a function of interaction strength and initial density. Thus, the effect is not limited to strictly one-dimensional systems. By going from a chain to the ladder, one can expect the existence of additional heavy excitations on the two-leg ladder that are defined on a rung, inherited from the $t_\perp \gg t_\parallel$ limit. Such objects can slow down the expansion even at low densities \cite{Hauschild2015} and in our investigation, we find that such additional heavy objects (or in other words, bound states) render the distillation process slower on ladders. Moreover, for both chains and ladders, the largest relative increase of the double occupancy in the center is obtained for {\it small} initial densities, whereas the purification of a clean band insulator requires densities to be close to $n_r\lesssim2$ to begin with \cite{Heidrich-Meisner2009}. Thus, initial densities close to half filling seem optimum in order to observe a large effect while still having a sizable amount of doublons in the system.

Second, we consider initial states with only one defect in the confined region (i.e., only one site where the density deviates from $n_r =2$) and compare the evaporation dynamics of singlons and holons on isotropic ladders and chains. While singlons escape via the quantum distillation mechanism, a single holon can only move once a doublon has partly dissolved into singlons or has slowly propagated as a whole, both of which happens on much slower time scales. Therefore, the density of holons in the initial state is the main limiting factor for the quantum distillation on quasi-one dimensional structures. Importantly, though, this does not lead to a bottleneck for the escape of singlons since they can exchange their position with both doublons and holons, and thus holons primarily reduce the achievable core density. The main difference between ladders and chains is traced back to the extra possibility of singlons to oscillate between the two legs as they escape and to a partial selftrapping of the singlon defect.

Third, many experiments with ultra-cold quantum gases in optical lattices start from product states rather than ground states in the initial trap \cite{Schneider2012,Trotzky2012,Ronzheimer2013,Schreiber2015}. While this case has been studied for bosons \cite{Muth2012}, no systematic study of this experimentally relevant initial condition has been carried out for fermions. We analyze product states with various concentrations of doublons, singlons, and holons as well as random or translationally invariant states. Concerning the efficiency, our results indicate that initial product states with a small number of (or without any) holons are ideal candidates for a fast and efficient quantum distillation. Curiously, the quantum distillation from engineered product states can be more efficient than from correlated initial states with the same average density. 

Clearly, an analysis of the most interesting regime of two-dimensional quantum gases would require an experimental effort, given the scarce set of available theoretical tools for the strongly interacting regime (see, e.g., the discussion in Ref.~\cite{Schluenzen2017}). Few-leg ladders have been realized in many ultra-cold quantum gas experiments using either supperlattices \cite{Atala2014}, digital mirror devices \cite{Tai2017} or synthetic lattice dimensions \cite{Mancini2015,Stuhl2015,Kolkowitz2017,Livi2016,Alex2017}. Note also that the ladders have been widely studied as models with surprising quantum properties such as spin gaps and superconductivity upon doping in condensed matter physics as well \cite{Dagotto1996}, with numerous realizations in quantum magnets \cite{Dagotto1999}.

The paper is organized as follows: In Sec.~\ref{sec:system} we present the model and the initial state preparation. Furthermore, we introduce the quantities investigated throughout our work. Section~\ref{sec:distill} is devoted to the presentation of the quantum distillation process on ladders. We focus on various system parameters with a detailed comparison between chains and two-leg ladders. In Sec.~\ref{sec:imp}, we analyze the defect evaporation, i.e., single singlon and holon dynamics in the doublon background. In Sec.~\ref{sec:prod}, we compare the expansion from product states to the expansion from correlated initial states. Finally, in Sec.~\ref{sec:conclusions}, we summarize our results. An appendix contains the calculation of the two-body spectrum on a two-leg ladder.

\section{Hamiltonian and setup}
\label{sec:system}

\begin{figure}[!t]
\includegraphics[width=1.0\columnwidth]{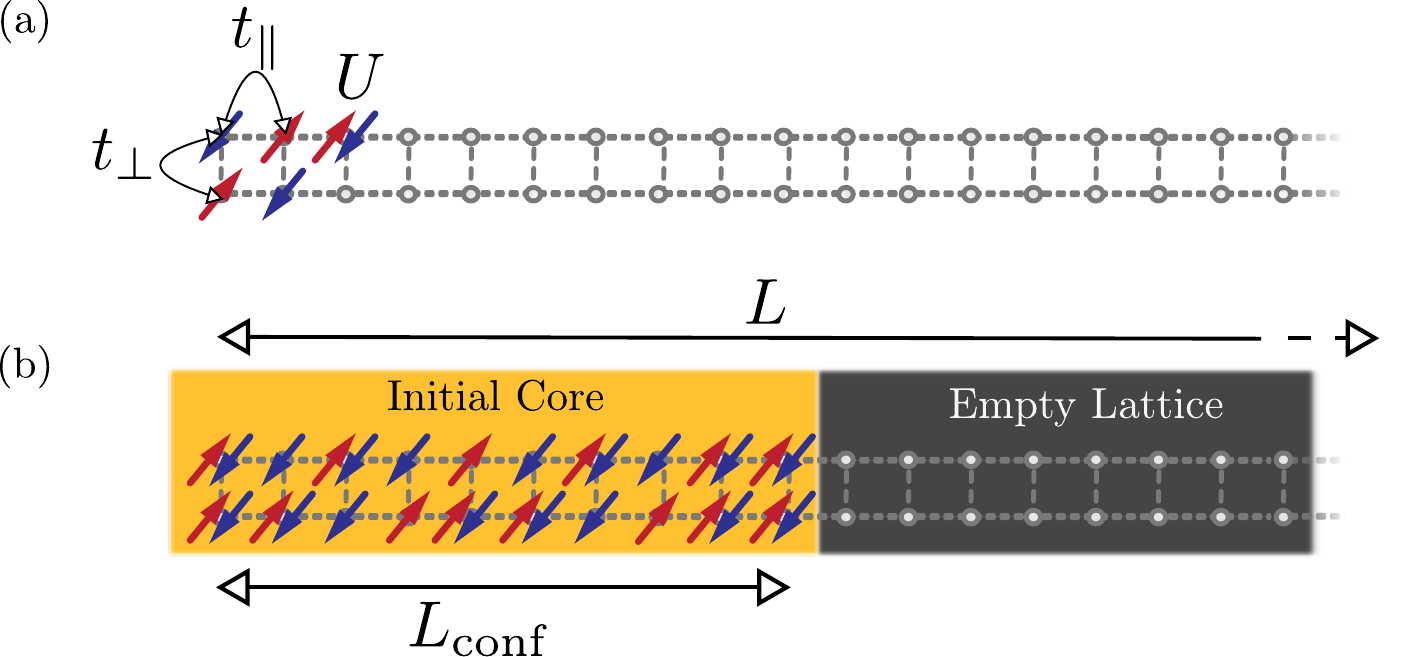}
\caption{(Color online) Schematic representation of (a) a two-leg ladder Hubbard model and (b) a typical initial state considered in this work (see the text for details).}
\label{fig2}
\end{figure}

We consider the Fermi-Hubbard model on the quasi-1D two-leg ladder geometry [see also the sketch in Fig.~\ref{fig2}(a)]
\begin{eqnarray}
H=&-&t_{\parallel}\,\sum_{r,\ell,\sigma}\left(
c^{\dagger}_{r,\ell,\sigma}c^{\phantom{\dagger}}_{r+1,\ell,\sigma}+\mathrm{H.c.}\right)\nonumber\\
&-&t_\perp\,\sum_{r,\ell,\sigma}\left(
c^{\dagger}_{r,\ell,\sigma}c^{\phantom{\dagger}}_{r,\ell+1,\sigma}+\mathrm{H.c.}\right)\nonumber\\
&+&U\,\sum_{r,\ell}n_{r,\ell,\uparrow}n_{r,\ell,\downarrow}\,,
\label{ham}
\end{eqnarray}
where $t_{\parallel}$ ($t_{\perp}$) is the leg (rung) hopping matrix element, $U$ is the on-site interaction, $c^\dagger_{r,\ell,\sigma}$ creates a fermion on the $r$-th rung of the $\ell$-th leg with spin $\sigma=\{\uparrow,\downarrow\}$, and $n_{r,\ell,\sigma}=c^\dagger_{r,\ell,\sigma}c^{\phantom{\dagger}}_{r,\ell,\sigma}$ is the density of fermions with spin $\sigma$ on $(r,\ell)$ site. The sum over $r$ goes over all $L$ sites within each leg, where the sum over $\ell$ goes over all sites in each rung. Furthermore, we use $t_{\parallel}=1$ (with $\hbar=1$). In the following we express time $\tau$  in dimensionless units, i.e., $\tau\,t_{\parallel}$, since $[\tau]=[1/t_\parallel]$. Also, we set lattice spacing to unity $a=1$. As a consequence the density quantities in this work are also dimensionless. Finally, for completeness, we define (i) the rung particle density as $n_{r}=\sum_{\ell,\sigma}\langle n_{r,\ell,\sigma}\rangle$ (with $\langle \cdot \rangle$ denoting the expectation value evaluated in a many-body state $|\psi \rangle$), (ii) the doublon $c^\dagger_{r,\ell,\downarrow}c^\dagger_{r,\ell,\uparrow}|0\rangle$ ($|0\rangle$ empty lattice site) with the associated doublon density $d_{r}=\sum_\ell d_{r,\ell}=\sum_\ell \langle n_{r,\ell,\downarrow}n_{r,\ell,\uparrow}\rangle$, and (iii) a singlon $c^\dagger_{r,\ell,\sigma}|0\rangle$ with the singlon density $s_{r}=n_{r}-2d_{r}$. In the rest of the paper, in order to facilitate a direct comparison between the chain and ladder geometry, we refer to the former as the sum of the uncoupled legs at $t_{\perp}/t_{\parallel}=0$. Furthermore, unless stated differently, we always quote the total density on a rung.

Unless stated otherwise, our initial states are prepared as the ground state of $H'=H+H_{\mathrm{conf}}$ for $U/t_{\parallel}=U_{\mathrm{GS}}/t_{\parallel}=0$, where
\begin{equation}
H_{\mathrm{conf}}=\epsilon\sum_{r=L_{\mathrm{conf}}+1}^{L}\sum_{\ell,\sigma}n_{r,\ell,\sigma}\,
\label{hamconf}
\end{equation}
is the confinement potential. Here, $\epsilon/t_{\parallel}\simeq10^4$ and $L_{\mathrm{conf}}$ is the number of rungs in the confined region (to which we will refer to also as the {\it core}). In order to decrease finite--size effects we place the core on the leftmost side of the system, see Fig.~\ref{fig2}(b) (in contrast to Fig.~\ref{fig1}, where we place it in the center of the lattice). At $\tau=0$, we suddenly switch off  the confinement potential ($\epsilon\to0$) and let the system evolve under the Hamiltonian \eqref{ham}  with the desired interaction strength $U$. In our setup the expansion will be asymmetric and the trapped gas will melt at the right edge of the confined region during the time evolution. It is worth noting that such a choice of the confinement potential does not change the general behaviour, as was shown in Refs.~\cite{Heidrich-Meisner2008,Heidrich-Meisner2009}. If not stayed otherwise, in this work we will use the two-leg ladder geometry ($\ell=1,2$) with $L=40$ rungs and set confined region to $L_{\mathrm{conf}}=10$ rungs. Furthermore, we impose open boundary conditions.

The nonequilibrium time evolution is studied by means of the time-dependent density-matrix renormalization group \cite{Vidal2004,White2004,Daley2004} (tDMRG) method with a third-order Trotter-Suzuki scheme. The specific implementation follows Ref.~\cite{White2004,Daley2004}, while a comprehensive introduction to the DMRG method can be found in Ref.~\cite{Schollwoeck2011,Feiguin2011,Feiguin2013a}. The premise is to obtain a wave-function that approximates the actual ground-state -- or time-evolved state -- in a reduced Hilbert space. The proposed solution has the very peculiar form of a matrix-product state, where the coefficients of the wave-function are obtained by contracting a product of matrices. The matrices are determined variationally, and the DMRG method is one way to do it efficiently. The accuracy of the wave function is typically quantified in terms of the discarded weight or truncation error, which decreases with the size of the matrices, or number of states kept (also called bond dimension $m$). The solution can be made asymptotically exact as this bond dimension approaches the total number of degrees of freedom. DMRG is formulated via diagonalizing reduced density matrices $\rho_A$ that are obtained by cutting the system into two parts $A$ and $B$ and then computing $\rho_A = \mbox{tr}_B |\psi\rangle \langle \psi |$, where $|\psi \rangle$ is the target wave-function. Diagonalization of $\rho_A$ leads to $\rho_A = \sum_{\alpha=1}^{s}  w_\alpha |\alpha \rangle \langle  \alpha |$ (where $s$ is the smaller of the Hilbert space dimensions of part $A$ and $B$). The approximation used is to truncate in the spectrum of $\rho_A$ and to keep only the $m\ll s$ states with the largest eigenvalues $w_1 > \dots w_m > \dots $.

During the time-dependent simulations, we allow the number of states to grow (controlled by a set discarded weight $\delta \rho = \sum_{\alpha>m }^s w_\alpha$) up to $m=2048$ states.  The time propagation uses the schemes described in Ref.~\cite{White2004,Daley2004} with a  time step of $\delta \tau\, t_{\parallel}=0.05$.  We carried out simulations with other time steps and also different discarded weights to ensure that the data are numerically accurate. At times $\tau t_\parallel \sim 30 $, the fastest particles have reached the boundary of an $L=40$ chain with $L_{\mathrm{conf}}=10$  and have propagated back to the position of the original interface between occupied and empty sites. This defines the largest time beyond which the escape dynamics inside the core $r \leq L_{\rm conf}=10$ can become system-size dependent.

A qualitative way to assess the accessible times in tDMRG simulations is to look at the time dependence of the entanglement entropy $S_{\rm vN} =-\mathrm{Tr}[ \rho_A \mathrm{ln} \rho_A ]$ (see the discussion in Ref.~\cite{Schollwoeck2011}). The behavior of the entanglement entropy for the quantum distillation was discussed for the Hubbard chain in much detail in Ref.~\cite{Heidrich-Meisner2009} and we did not observe noticeable differences in the case of ladders. Among time-evolution problems, the sudden expansion is a more benign problem than, e.g.,  global quenches (where $S_{\rm vN}\propto t $) due to the inhomogeneity of how $S_{\rm vN}$ grows and its overall slower increase. The time dependence in related geometric quenches was studied in great detail in, e.g., Ref.~\cite{Alba2014}.

\section{Results}
\label{sec:distill}

\subsection{Quantum distillation on isotropic ladders}

\begin{figure}[!t]
\includegraphics[width=1.0\columnwidth]{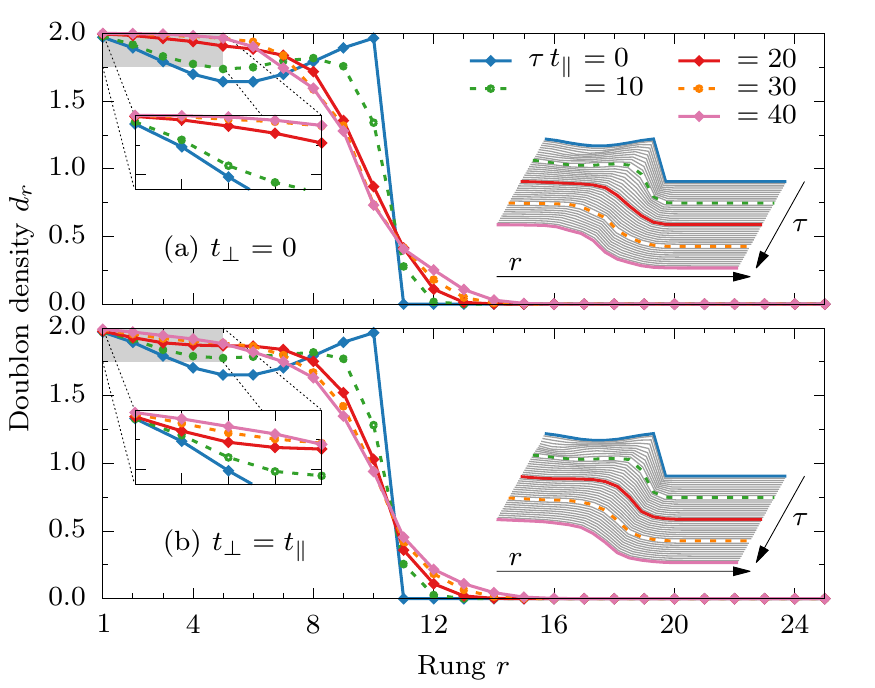}
\caption{(Color online) Snapshots of the rung doublon density $d_{r}$ for (a) uncoupled chains ($t_\perp/t_{\parallel}=0$) and (b) the isotropic ladder ($t_\perp/t_{\parallel}=1$) calculated for an initial density $n_{\mathrm{conf}}=1.9$ in the confined region ($L_{\mathrm{conf}}=10$, $N=38$), and interaction strength $U/t_{\parallel}=40$. Insets in (a) and (b): (left) zoom on $d_{r}$ profiles of the $5$ leftmost rungs, and (right) snapshots of $d_{r}$ for $r=1,\dots,20$ and $\tau\,t_{\parallel}=0,\dots,40$.}
\label{fig3}
\end{figure}

\begin{figure*}[!t]
\includegraphics[width=1.0\textwidth]{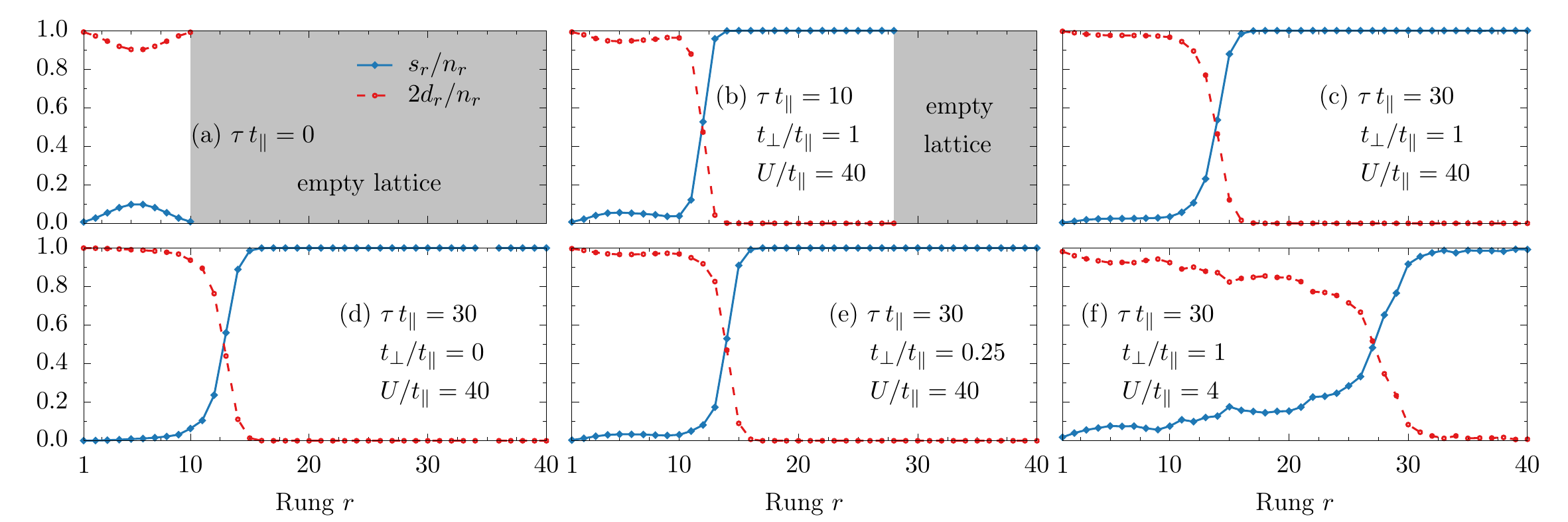}
\caption{(Color online) Dynamical spatial separation of singlons and doublons as calculated for $L=40$, $L_{\mathrm{conf}}=10$, and $n_{\mathrm{conf}}=1.9$. Note that here we plot the {\it relative} singlon and doublon densities $s_r/n_r$ and $2d_r/n_r$, respectively. (a) Initial state, $\tau\,t_{\parallel}=0$, for all configurations studied. (b-c) Spatial singlon-doublon separation for $U/t_\parallel=40$ and isotropic ladder $t_{\perp}/t_{\parallel}=1$ [for the same parameters as in Fig.~\ref{fig3}(b)] for (b) $\tau\,t_{\parallel}=10$ and (c) $\tau\,t_{\parallel}=30$. (d-e) Spatial separation for $U/t_{\parallel}=40$, (d) uncoupled chains $t_{\perp}/t_{\parallel}=0$, and (e) a weak rung tunneling $t_{\perp}/t_{\parallel}=0.25$. In panel (e), we present evidence for the disappearance of the dynamical separation in the core for the case of the isotropic ladder at small interaction strength $U/t_{\parallel}=4$.
}
\label{fig4}
\end{figure*}

In Fig.~\ref{fig3}, we present a comparison of snapshots of the rung double occupancy $d_{r}$ for uncoupled chains ($t_{\perp}/t_{\parallel}=0$) and for the isotropic ladder ($t_{\perp}/t_{\parallel}=1$) at $U/t_\parallel=40$. Both systems exhibit a similar melting dynamics at the interface of the initially occupied region and the empty lattice. Furthermore, in both cases we observe an increase of the doublon density beyond its initial value in the core of the system. It is thus evident that the strong version of quantum distillation takes place in the ladder geometry. A similar behavior is also observed in the case of a three--leg ladder on smaller lattice $L$ (not shown).

Regarding the time scales, in the related sudden expansion experiments, the dynamics can be recorded up to an increase of the initial cloud size by a factor of two for fermions \cite{Schneider2012} or four to five for bosons \cite{Ronzheimer2013}. This growth is covered in our simulations as well, yet for smaller particle numbers per 1D system than what can typically be realized in these experiments.

The data presented in Fig.~\ref{fig3} shows that the initial density for correlated initial states is typically nonuniform, due to Friedel oscillations induced at the boundaries. The averaging-out of these inhomogeneities after the removal of the boundary naturally leads to a transient increase in the middle of the initially occupied region (i.e., $r=L_{\rm conf}/2$) but is not the reason for the strong quantum distillation, which manifests itself in an increase on all sites  $r \leq L_{\rm conf}/2$ for the parameters
of the figure.

In order to further corroborate our observations, we next discuss the dynamical spatial separation of singlons and doublons during the quantum distillation process on the ladder. In Fig.~\ref{fig4}, we present the density of singlons $s_r$ and doublons $d_r$ on each rung. Note that we normalize each of the densities to the total particle density $n_r$ on the same rung, i.e., $s_r/n_r$ and $2d_r/n_r$. These normalized densities are equal to $1$ if a singlon (doublon) is the lone particle on a given rung. In Fig.~\ref{fig4}(a), we present our initial state, $\tau\,t_{\parallel}=0$, where all particles are present only in the core, $r \le L_{\mathrm{conf}}$, and the rest of the lattice is empty. Figure~\ref{fig4}(b) illustrates the behavior at a transient time $\tau\,t_{\parallel}=10$, for which we can observe that only singlons (which escaped from the core) propagate in the empty lattice [see also Fig.~\ref{fig1}(c)]. For the largest simulation time reached, before reflections off the far boundary start to matter [$\tau\,t_{\parallel}=30$, see Fig.~\ref{fig4}(c)], all the rungs (sites) outside the core are occupied solely by singlons. On the other hand, it is clear from the presented results that the doublons remain in the core. Note that we observe the same behavior also for the case of uncoupled chains, see Fig.~\ref{fig4}(d).

Although the overall behaviour in chains and ladders is akin, there are also some differences in the dynamics of these two setups. While for the 1D system \cite{Heidrich-Meisner2009} (or uncoupled chains) the density in the first rungs increases up to its maximal value of $n_r=2$, thus forming a band insulator, for the ladder geometry, this maximum occupation is never reached at this value of $U/t_{\parallel}$ and $n_{\rm conf}=1.9$ [see the insets of Fig.~\ref{fig3}(b)], at least on the time scales of our simulations. Thus, as expected, on ladders, the quantum distillation is somewhat less efficient than in strictly 1D systems. One reason is the possibility of a singlon to just keep exchanging its position with a doublon on the same rung. Another origin of slower dynamics on the ladder is a partial selftrapping of singlons. We will discuss both of these scenarios in more detail in Sec.~\ref{sec:imp}.

\begin{figure}[!t]
\includegraphics[width=1.0\columnwidth]{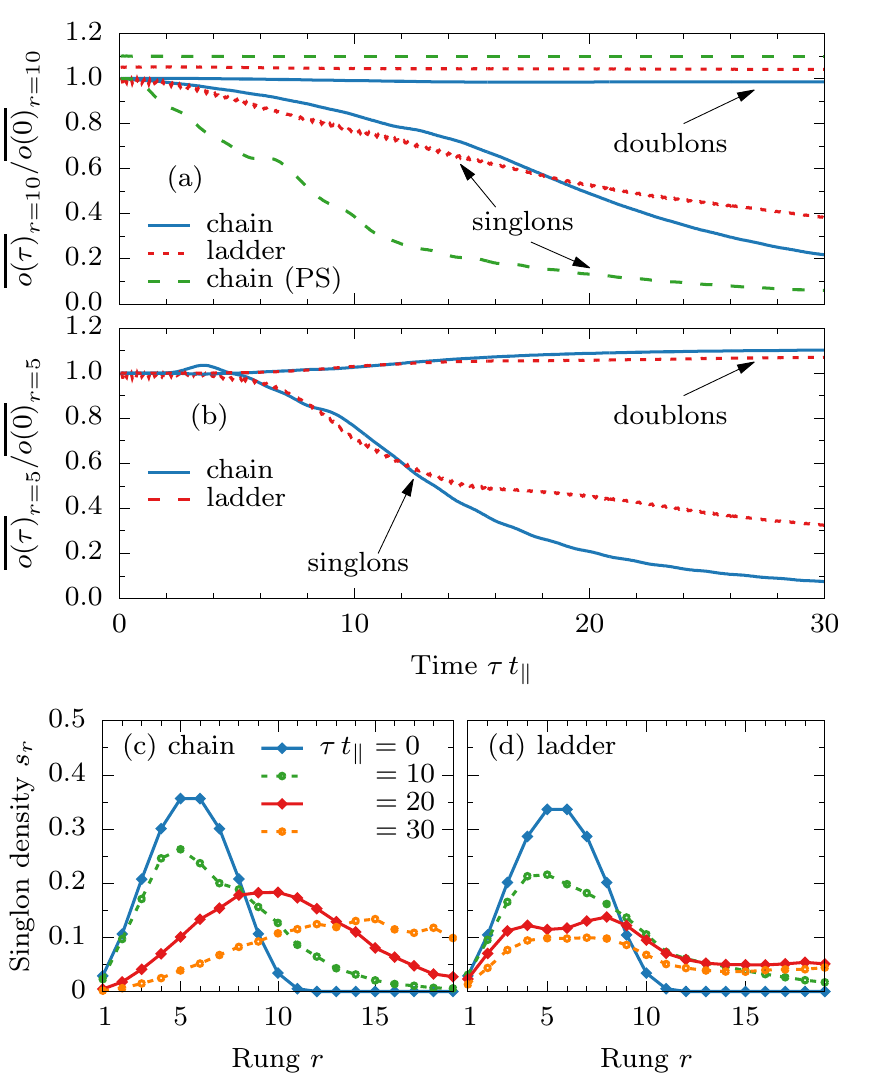}
\caption{(Color online) Time dependence of the average doublon and singlon occupancy averaged over the first (a) $r=10$ and (b) $r=5$ rungs as calculated for $U/t_{\parallel}=40$ and $n_{\mathrm{conf}}=1.9$. In addition, in  (a) we show results obtained from an initial product state (PS) [system parameters are the same as in Fig.~\ref{fig12}(i); see Sec..~\ref{sec:prod} for details]. Note that in (a), results for the ladder and for the expansion from a product state on the chain are shifted by $0.05$ and $0.1$, respectively. (c,d) Snapshots of the singlon density profiles for (c) the chain and (d) the ladder geometry and at different times (see the legend). The model parameters are the same as in Fig.~\ref{fig3}.}
\label{fig5}
\end{figure}

In order to investigate the quantum distillation on ladders in more detail, we analyze the time evolution of the average doublon and singlon densities on the first $r$ rungs,
\begin{equation}
\overline{s}_{r}=\frac{1}{r}\sum_{i=1}^{r}s_{i}\,,\qquad
\overline{d}_{r}=\frac{1}{r}\sum_{i=1}^{r}d_{i}\,.
\label{eq:denav}
\end{equation}
In Fig.~\ref{fig5}(a,b) we present results for $r=L_{\mathrm{conf}}=10$ and $r=5$. As is clearly visible in Fig.~\ref{fig5}(a), the average doublon density on the first $L_{\mathrm{conf}}$ sites remains constant, as expected. Simultaneously, the singlon density steadily decreases. Such results are again indicative of the presence of the distillation process. However, in order to distinguish between the weak or strong version of the latter it is better to measure the average density of part of the core. In  Fig.~\ref{fig5}(b) we present results for $r=5$ where for both types of lattices we see not only a decrease of the number of singlons, but also an increase in the doublon density of the core. While, for both chain and ladder lattice geometry we observe the strong version of distillation, the internal times scales are different. Namely, for $\tau\,t_{\parallel}\lesssim10$, the dynamics of both systems is essentially the same. Next, for larger times, we observe that singlons {\it escape} the core faster for uncoupled chains. For $\tau\,t_{\parallel}\gg 1$, the number of singlons decreases to a small fraction of its initial value. At $\tau/t_{\parallel}=30$, only $\sim5\%$ of the singlons remain in the core for the chain and about $\sim20\%$ for the isotropic ladder. This is clearly visible in Figs.~\ref{fig5}(c) and (d), where we present snapshots of the singlon density at different times and for the same parameters as in Fig.~\ref{fig3}.

\subsection{Dependence on model parameters}

\begin{figure}[!t]
\includegraphics[width=1.0\columnwidth]{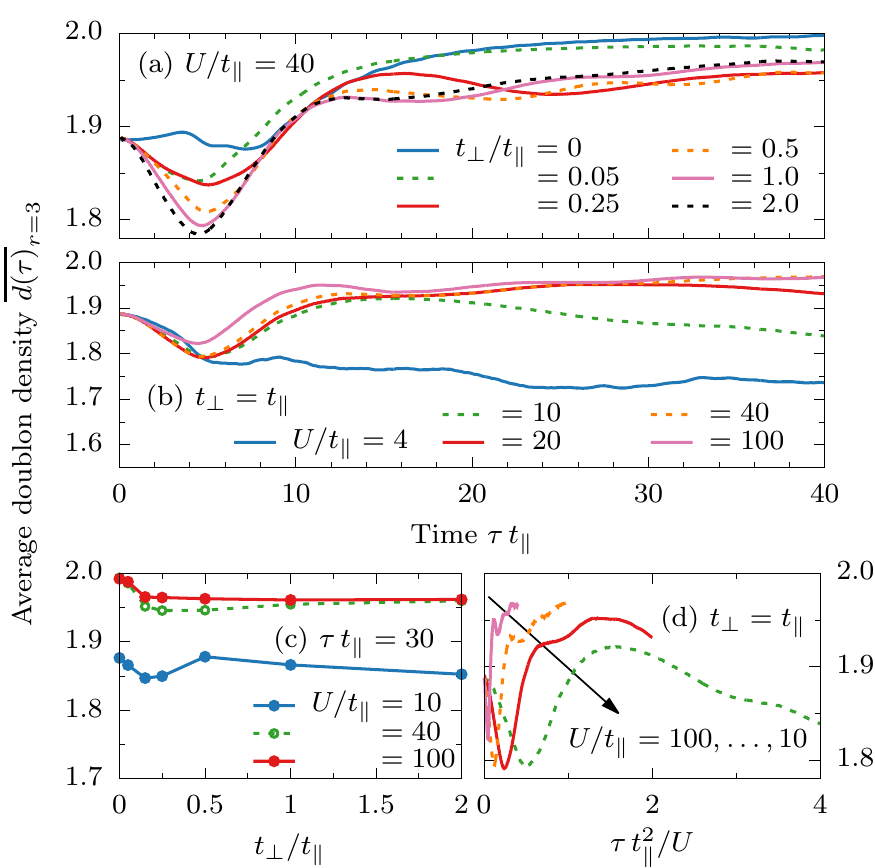}
\caption{(Color online) Time dependence of average rung double occupancy $\overline{d}_{r}$ for $r=3$: (a) initial confinement density $n_{\mathrm{conf}}=1.9$, and interaction $U/t_{\parallel}=40$ for various $t_\perp/t_{\parallel}$; (b) $n_{\mathrm{conf}}=1.9$, isotropic ladder for various $U/t_{\parallel}$. (c) $t_\perp$-dependence of $\overline{d}_{r=3}$ for $\tau\,t_{\parallel}=30$, $n_{\mathrm{conf}}=1.9$, and various $U/t_{\parallel}$. Note that this result does not reflect the $\tau\sim \tau_{\rm QD}$ limit (see the text for details). (d) $\overline{d}_{r=3}$ as a function of $\tau t_{\parallel}^2/U$ for $n_{\mathrm{conf}}=1.9$ and $t_\perp/t_{\parallel}=1$.}
\label{fig6}
\end{figure}

Let us now focus on the dependence of the distillation process on $t_\perp/t_{\parallel}$ and $U/t_{\parallel}$. First, we set $U/t_\parallel=40$ and vary $t_\perp$. Our results presented in Fig.~\ref{fig6}(a) indicate that for all values of $t_\perp$ considered here ($0\leq t_\perp/t_\parallel \leq 2$),  there is an increase of the core density beyond its initial value for $\tau t_{\parallel}>10$ after some transient drop. Thus, although quantitatively less efficient than at $t_{\perp}/t_{\parallel}=0$, the strong version of quantum distillation takes place. Interestingly, the core density of doublons resulting from the quantum distillation is independent of the strength of the rung-tunneling matrix element already for $t_{\perp}/t_{\parallel}\gtrsim0.2$ [see Fig.~\ref{fig6}(c)]. One needs to keep in mind, though, that the results presented in Fig.~\ref{fig6}(c) may not reflect the $\tau\sim \tau_{\rm QD}$ behavior yet, since for some values of $t_\perp$, the density still increases. Similarly, dynamical spatial separation of singlons and doublons exists for all values of $t_{\perp}/t_{\parallel}$. We illustrate this in Figs.~\ref{fig4}(d),(e) and (c) for $t_{\perp}/t_{\parallel}=0\,,0.25\,,1.0$, respectively. 

Second, we keep $t_\perp/t_{\parallel}=1$ fixed and vary $U/t_{\parallel}$. In Fig.~\ref{fig6}(b), we present the interaction dependence of our results. On grounds of energy conservation and of the requirement of the doublons to be long-lived objects, it is obvious that the lower bound on the interaction strength for quantum distillation to occur is $U_b\gtrsim W$, where $W$ is the bandwidth. In the case of the isotropic ladder, $W/t_{\parallel}=6$. Consistent with this qualitative argument, our numerical results suggest that the behavior is similar to the $t_{\perp}/t_{\parallel}=0$ case \cite{Heidrich-Meisner2009}, i.e., the process is the most efficient for $U/t_{\parallel}\gtrsim10$. Furthermore, we also find a universal time scale $\propto\tau t_{\parallel}^2/U$ [compare Figs.~\ref{fig6}(b) and (d)], where results for various values of the interaction strength $U$ have the same - initial-filling dependent - behavior at long times. This time scale captures the slow melting of the block of doublons that was formed through the quantum distillation process.

For $U/t_{\parallel}=4$, we observe that the average density of the doublons in the core decreases steadily with time. This indicates that doublons do not remain  anymore in the core but, simultaneously with the singlons, propagate through the lattice. Note, however, that singlons and holons move through the lattice at different speeds. While singlons propagate through the empty lattice with the maximal possible velocity, $v_{s}=2t_{\parallel}$, doublons move much slower. As a consequence, one still observes a dynamical separation of singlons and doublons at transient times. This is also presented in Fig.~\ref{fig4}(f), where for an interaction strength of $U/t_{\parallel}=4$ we do not observe singlon--doublon separation in the core anymore, while the interface between regions with mostly doublons versus a region with mostly singlons is now in the region where the lattice was initially empty.

\begin{figure}[!t]
\includegraphics[width=1.0\columnwidth]{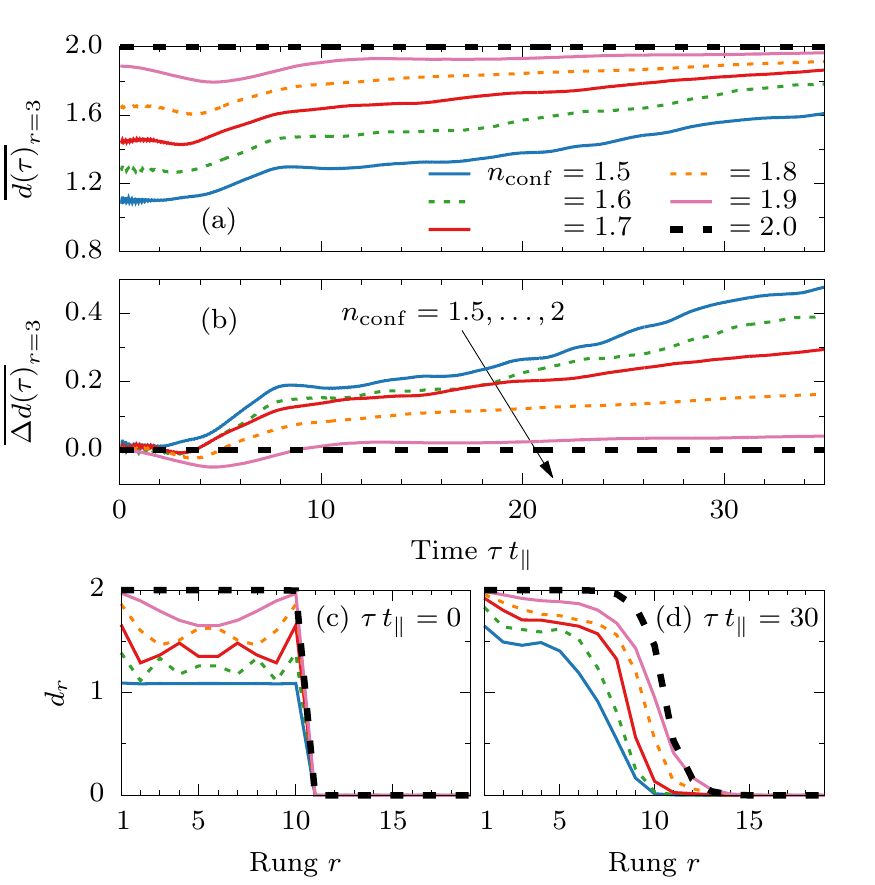}
\caption{(Color online) Time dependence of (a) the average rung doublon occupancy $\overline{d}_{r}$ and (b) the relative average rung doublon occupancy $\overline{\Delta d(\tau)}_r$ for the first $r=3$ sites, fixed $U/t_{\parallel}=40$, $t_{\perp}/t_{\parallel}=1$, and various initial $n_{\mathrm{conf}}$. (c,d) Density profiles $d_{r}$ for (c) $\tau\,t_{\parallel}=0$ and (d) $\tau\,t_{\parallel}=30$ for the same parameters as presented in panel (a). The legend of (a) applies to all panels.}
\label{fig7}
\end{figure}

Next, we comment on the dependence of the distillation on the initial filling $n_{\mathrm{conf}}$ in the confinement region. In Fig.~\ref{fig7}(a) we present the time dependence of the average doublon density in the first $r=3$ rungs of the core. As is clearly visible, for all considered values of $n_{\mathrm{conf}}$, we observe an increase of double occupancy. Moreover, our results indicate that for setups with small values of $n_{\mathrm{conf}}$, the processes of quantum distillation leads to a larger {\it relative increase} of the core double occupancy [see Fig.~\ref{fig7}(b), where we present $\overline{\Delta d(\tau)}_r=\overline{d(\tau)}_r/\overline{d}(0)_r-1$], related to the fact that in our initial states, prepared with $U_{\mathrm{GS}}/t_{\parallel}=0$, a small $n_{\mathrm{conf}}$ is a result of a large number of singlons. For instance, for a density of $n_{\mathrm{conf}}=1.5$ ($n_{\mathrm{conf}}=1.9$) we have $\sim10$ ($\sim18$) doublons and $\sim10$ ($\sim 2$) singlons in the initial state. 

\begin{figure}[!t]
\includegraphics[width=1.0\columnwidth]{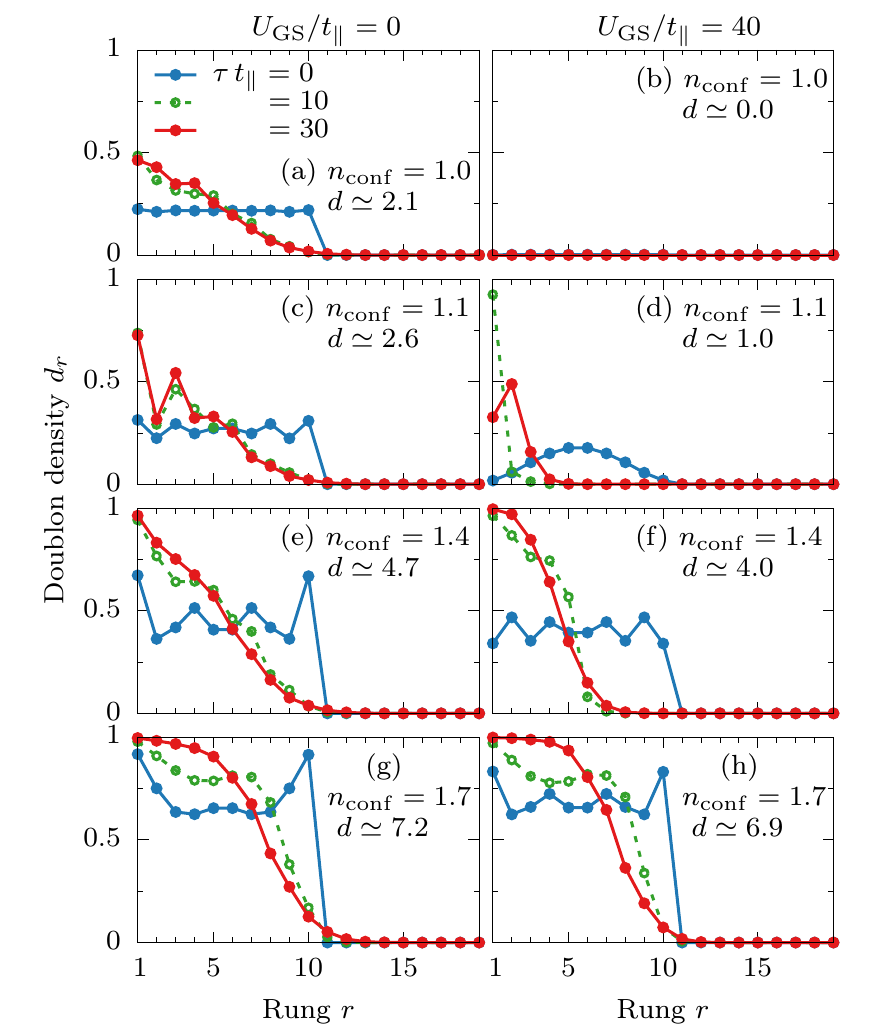}
\caption{(Color online) Snapshots of density profiles for various initial densities $n_{\rm conf}=1.0,1.1,1.4$, and $1.7$, obtained for the expansion from the ground state calculated with (a),(c),(e),(g) $U_{\mathrm{GS}}/t_{\parallel}=0$ and (b),(d),(f),(h) $U_{\mathrm{GS}}/t_{\parallel}=40$. $d$ is the total number of doublons in the system at time $\tau=0$.}
\label{fig7b}
\end{figure}

Finally, and to conclude this section, it is worth noting that the ratio between singlons and doublons in the initial state can also be controlled by the interaction $U_{\mathrm{GS}}/t_{\parallel}$ with which the initial state is obtained. Preparing the system in a correlated state with $U_{\mathrm{GS}}/t_{\parallel}\ne0$ leads to a decrease of the doublons-to-singlons ratio in comparison to a noninteracting gas with $U_{\mathrm{GS}}/t_{\parallel}=0$. As a consequence, the time evolution from  the latter initial state can exhibit a larger relative increase of the double occupancy. The results for $U_{\mathrm{GS}}/t_{\parallel}=0$ and $U_{\mathrm{GS}}/t_{\parallel}=40$ for a chain  presented in Fig.~\ref{fig7b} confirm such a behaviour. Within our core size $L_{\rm conf}$, for small $n_{\rm conf}\sim1$, $U_{\mathrm{GS}}/t_{\parallel}=0$ is the better choice for the initial state since $U\gg t_\parallel$ leads to a Mott insulator (for exactly $n_{\rm conf}= 1$). In the latter case, the doublon density is suppressed   and quantum distillation does not take place. On the other hand, for a modest value of $n_{\rm conf}\sim1.5$, the correlated state with a large doublons-to-singlons ratio leads to, again, a large relative increase of the double occupancy. Finally, effects of $U_{\mathrm{GS}}$ vanish in the $n_{\rm conf}\to2$ limit, since every site is almost doubly occupied.

\section{Expansion from Fock states}

In Sec.~\ref{sec:distill}, we chose as the initial state the ground state of our system in a box trap with $U_{\mathrm{GS}}/t_{\parallel}=0$. As a consequence, in the core, we have a superposition of singlons and doublons on every site [see Fig.~\ref{fig4}(a)]. However, many optical lattice experiments that study nonequilibrium dynamics \cite{Trotzky2012,Schneider2012,Ronzheimer2013,Pertot2014,Vidmar2015,Schreiber2015,Choi2016} start from product states rather than ground states, including some sudden-expansion experiments \cite{Schneider2012,Ronzheimer2013,Vidmar2015}. In order to account for this experimentally relevant condition, we next consider the time evolution starting from product states, i.e.,
\begin{equation}
|\psi_{\rm prod}\rangle = \prod_{i=1}^{L_{\rm conf}}
c_{i,\uparrow}^{n_{i,\uparrow}} c_{i,\downarrow}^{n_{i,\downarrow}}|0\rangle\,,
\end{equation}
where $n_{i,\sigma}=0,1$ are specially chosen integers. Thus, each site is occupied by exactly one singlon, doublon, or holon [for the ladder, $i=(r,\ell)$].

We will first study product states that have doublons on all but one or two sites to study the dynamics of individual defects in Sec.~\ref{sec:imp}.
Then we will consider more general classes of product states with lower average densities  in Sec.~\ref{sec:prod}.

\subsection{Defect evaporation}
\label{sec:imp}

In this section we investigate the behaviour of single {\it defects} in the background of doublons on the two-leg ladder geometry. In order to do so, we prepare an initial state with doublons on all but one or two sites in the confinement region, $r \leq L_{\mathrm{conf}}=10$. Such  vacancies are obtained with the help of large, local (one-site) potentials on the desired site(s). Singlons or holons are placed on sites in the first, leftmost rung $r=1$. On that rung, we consider: one singlon on one leg ($N=39$ particles in the core), two singlons, one on each leg ($N=38$ particles), one holon on one leg ($N=38$ particles), two holons, one on each leg ($N=36$ particles). 

Note that the initial placement of a given defect does not change the overall behaviour discussed in this section. Putting the vacancy somewhere {\it in the middle} of the core (i.e., for the setup presented in Fig.~\ref{fig1}) would lead to a propagation in two directions, each with properties as presented below. The quantum distillation happens regardless of where singlons initially sit in the initial state: remember that experiments use a block of particles that can melt to either side and both left- and right-moving portions of an initially localized defect can thus eventually leave the cluster, always resulting in doublons moving into the core. In our case of an asymmetric expansion, the portion of a singlon moving inside would  be reflected at the (left) boundary and then also eventually leave the cluster of doublons after traversing through it twice.

\subsubsection{Singlon defects}

\begin{figure}[!t]
\includegraphics[width=1.0\columnwidth]{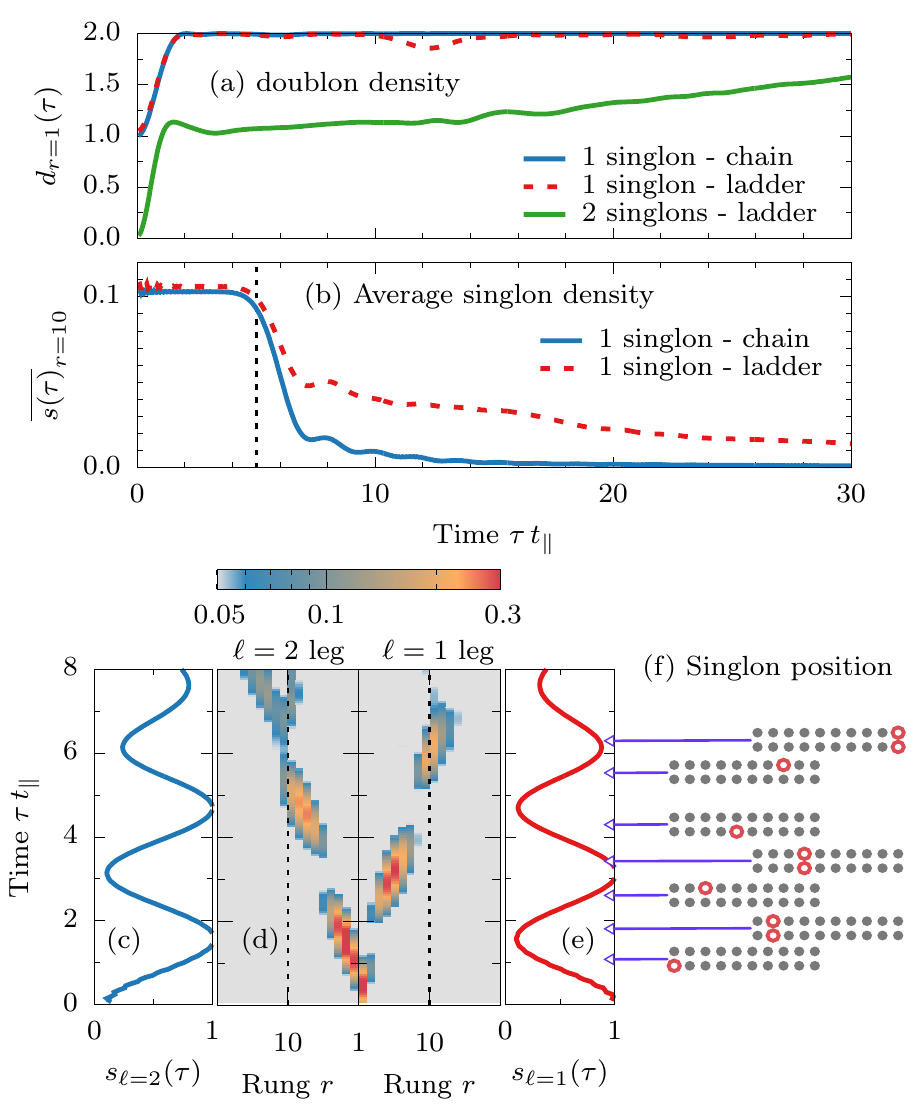}
\caption{(Color online) (a) Time dependence of the double occupancy $d_{r=1}$ on the first rung for initial states with a singlon placed on the first rung $r=1$. See the text for details on the initial state ($t_{\perp}/t_{\parallel}=0\,,1$, $L_{\mathrm{conf}}=10$, and $U/t_{\parallel}=40$). (b) Time evolution of the average singlon density in the initially confined region, $\overline{s(\tau)}_{r=10}$, for chain and isotropic ladder. The vertical dashed line indicates the time at which the singlon reaches the end of the core at $\tau\simeq L_{\mathrm{conf}}/v_s=5t_{\parallel}$. (d) Position- ($x$-axis) and time- ($y$-axis) dependence of the singlon density $\langle s_{r,\ell}\rangle$ within the isotropic ladder, $t_{\perp}/t_{\parallel}=1$, for $L=40$ rungs and $U/t_{\parallel}=40$. The boundary of the initially occupied region is depicted as a vertical dashed line. (c,e) Time evolution of the total singlon density $s_\ell$, Eq.~\eqref{eqsiru}, on the leg (c) $\ell=2$ and (e) $\ell=1$. (f) Schematic representation of the position of the singlon (red open circle) within the core at certain points in time: the data unveil that the singlon, while moving towards the edge, also oscillates between the two legs.}
\label{fig8}
\end{figure}

In Fig.~\ref{fig8}(a), we present the time evolution of the doublon density $d_{r=1}$ on the first rung for $U/t_{\parallel}=40$ for one and two singlons placed on the first, $r=1$, rung. Some general features can be inferred: (i) a single singlon escapes the initially occupied site with the same speed for chains and isotropic ladders. (ii) The wavefront of a single singlon reaches the end of the core ($r=L_{\rm conf}$) at time $\tau=L_{\mathrm{conf}}/v_s=5$ for both lattices [depicted also in Fig.~\ref{fig8}(b) as a vertical dashed line]. This represents the fastest possible propagation with the velocity $v=2t_{\parallel}$. (iii) The fast transport of singlons breaks down if two singlons are placed next to each other. In Fig.~\ref{fig8}(a), we present the doublon density on the rung where two singlons were initially placed (one on each leg). Initially, $d_{r=1}$ rapidly increases in a similar manner as in the case of a single singlon. At $\tau\,t_{\parallel}\sim 2$, the doublon density reaches $d_{r=1}\simeq1$, indicating that one singlon escaped the first site. Subsequently, for $\tau\,t_{\parallel}>2$, the doublon density increases very slowly and, as a consequence, the second singlon takes much longer to escape from that rung. We argue that this can be viewed as a partial selftrapping of singlons due to bound states, to be discussed in more detail below.

\begin{figure}[!t]
\includegraphics[width=1.0\columnwidth]{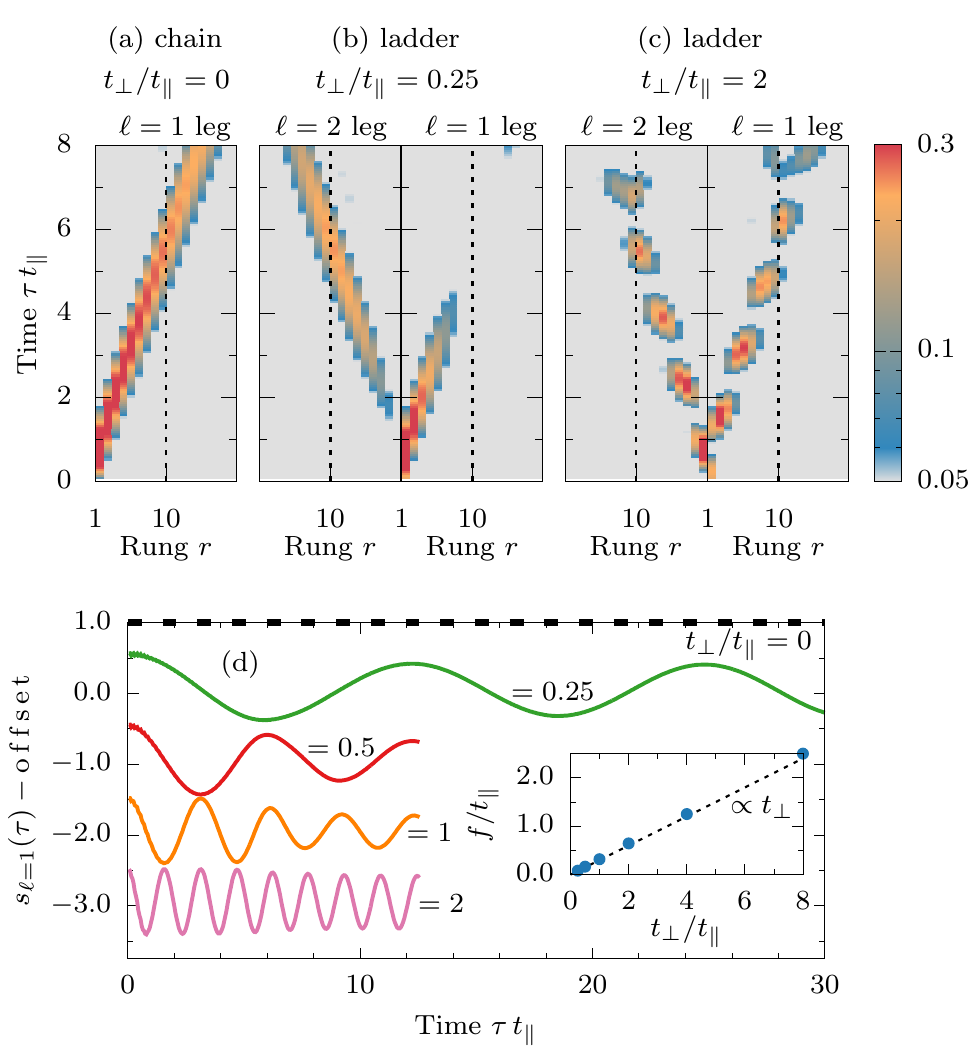}
\caption{(Color online) (a-c) Position- ($x$-axis) and time- ($y$-axis) dependence of the singlon density $\langle s_{r,\ell}\rangle$ for (a) $t_{\perp}/t_{\parallel}=0$, (b) $t_{\perp}/t_{\parallel}=0.25$, and (c) $t_{\perp}/t_{\parallel}=2$ for both legs of a ladder with $U/t_{\parallel}=40$. (d) Time evolution of the total singlon density on§ the leg where the singlon was initially placed, i.e., $s_{\ell=1}$, for $t_{\perp}/t_{\parallel}=0,0.05,0.25,0.5,1,2$ (top to bottom). For clarity, the results for $t_{\perp}/t_{\parallel}=0.25,0.5,1,2$ are shifted by $-0.5,-1.5,-2.5,-3.5$, respectively. Inset: rung--tunneling dependence of the frequency $f=1/\Delta\tau_\mathrm{osc}$ of singlon-density oscillations. The dashed line represents the $f\propto t_{\perp}$-dependence.}
\label{fig9}
\end{figure}

Although a single singlon escapes the first rung in a similar manner for the chain and isotropic ladder, the overall dynamics of singlons in the sea of doublons crucially depends on the lattice geometry. In Fig.~\ref{fig8}(b), we present the time dependence of the average number of singlons in the core $\overline{s(\tau)}_{r=10}$. Initially, there is no time dependence, reflecting that the singlon still propagates within the core. The wavefront reaches the right end of the occupied region at time $\tau\,t_{\parallel}=5$ for both lattices. However, the remaining fraction of the singlon escapes the core with a different time dependence. For the chain, the core density of singlons drops rapidly for $\tau\,\,t_{\parallel}>5$ and reaches its minimum, $\sim 0$, at $\tau\,t_{\parallel}\simeq15$. On the other hand, in the case of the ladder, $\overline{s}_{r=10}$ linearly decreases with time, reaching, at our largest simulation time $\tau=30t_{\parallel}$, $\sim10\%$ of its initial value. Note that a similar behaviour is also observed in Fig.~\ref{fig5}(b), where we present the time dependence of the density of the escaping singlons when the initial state is the ground state of the trapped gas with $U_{\mathrm{GS}}/t_{\parallel}=0$.

To gather further insight into this dichotomy, in Fig.~\ref{fig8}(d) we investigate the position of most of the singlon density, $\langle s_{r,\ell}\rangle>0.05$ [singlon density on a given site $(r,\ell)$], within both legs, $\ell=1,2$, on the isotropic two-leg ladder geometry. Initially, for $\tau\,t_{\parallel}<0.5$, the singlon expands only on the leg on which it was placed, i.e., the $\ell=1$ leg. Next, until time $\tau\,t_{\parallel}\simeq 1$, most of the particle is transferred to the other leg. Remarkably, as is clearly visible in Fig.~\ref{fig8}(d), for the next $\Delta\tau_\mathrm{osc}\simeq 1/t_{\perp}$, singlons propagate solely on the other leg ($\ell=2$). Subsequently, the whole process repeats and between $2.5\lesssim \tau\,t_{\parallel}\lesssim3.5$, the propagation takes place again only in the first $\ell=1$ leg. Such {\it oscillations} are clearly visible in the total singlon density of a given leg $\ell$,
\begin{equation}
s_{\ell}=\sum_{r=1}^{L}\langle s_{r,\ell}\rangle\,,
\label{eqsiru}
\end{equation}
presented in Figs.~\ref{fig8}(c) and (e). Note that $s_{\ell=2}(\tau)=s_\mathrm{ini}-s_{\ell=1}(\tau)$, where $s_\mathrm{ini}$ is the initial density of the singlons (in the case discussed in this section, $s_\mathrm{ini}=1$). The time evolution of $s_\ell$ yields information about the frequency $f$ with which singlons "hop" between the legs. Such a frequency can be determined from the time $\Delta\tau_\mathrm{osc}=1/f$ between two consecutive wave crests in the time dependence of the total singlon density in a given leg, i.e., $s_{\ell=1}(\tau)$. In Fig.~\ref{fig9}, we present the $t_{\perp}$-dependence of the singlon density $\langle s_{r,\ell}\rangle$ on all sites [panels (a-c)] together with the singlon density $s_{\ell=1}$ on the $\ell=1$ leg [panel (d)] as a function of time. It is evident from the latter [see the inset of Fig.~\ref{fig9}(d)] that $f=\alpha\,t_{\perp}$, where $\alpha$ is some constant.

\begin{figure}[!t]
\includegraphics[width=1.0\columnwidth]{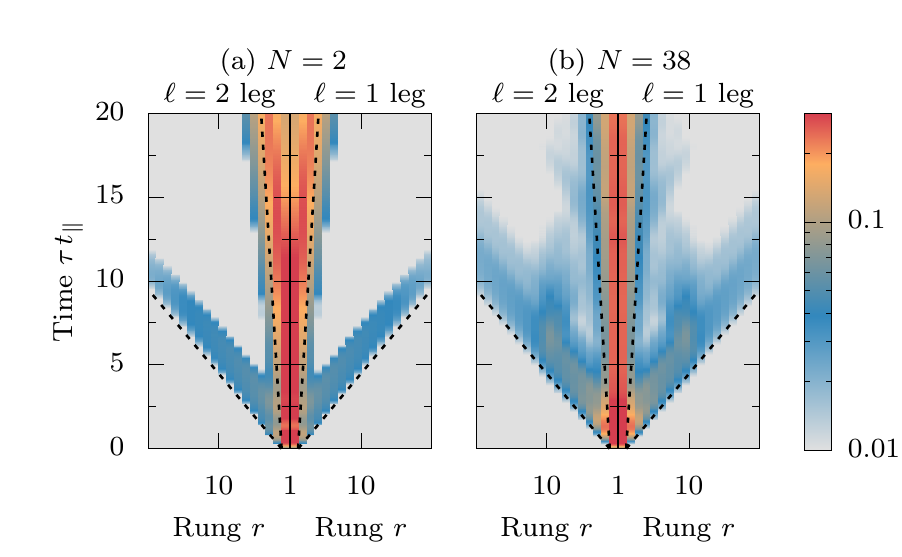}
\caption{(Color online) Position- ($x$-axis) and time- ($y$-axis) dependence of the singlon density $\langle s_{r,\ell}\rangle$ for two singlons placed on the first rung ($r=1$). (a) Results for two singlons in an otherwise empty lattice for $t_{\perp}/t_{\parallel}=2$, $U/t_{\parallel}=40$, and $N=2$. (b) Results for two singlons placed in the sea of doublons for $t_{\perp}/t_{\parallel}=1$, $L_{\mathrm{conf}}=10$, $U/t_{\parallel}=40$, and $N=38$ [compiled from the corresponding data for two singlons presented in Fig.~\ref{fig8}(a)]. The dashed lines in both panels represent {\it light-cones}, i.e., the fastest velocities of singlons ($v/t_{\parallel}=2$) and heavy objects ($v/t_{\parallel}=0.23$).}
\label{fig10}
\end{figure}

The selftrapping, which requires more than one singlon to be present in the initial state, can be understood from the strong-rung limit $t_\perp \gg t_\parallel$. The argument relies on a particle-hole transformation applied to the  initially occupied region: if there is one singlon (i.e., a hole on the doublon background), then this is equivalent to just a free particle in an empty system, while two singlons in a background of doublons correspond to two particles, which can have additional bound states originating from the $t_\perp \gg t_\parallel$ limit. Imagine isolated rungs which have single-particle energies $\epsilon_{\pm}= \pm t_\perp$. Thus, moving one singlon from a rung into a neighboring empty one is a resonant process for an infinitesimally small $t_\parallel$. If there are two singlons on a rung, then the two-particle energies are $\epsilon_{2s} = -2t_\perp, 0,0, 2t_\perp$. If the initial state has an overlap with the local $\epsilon_{2s}=0$ states§, then this is off-resonant with two singlons in two rungs in the $\epsilon_{-}=-t_\perp$ state, while moving one into $\epsilon_{-} = -t_\perp $ and another one into $\epsilon_{+} = t_\perp$ remains possible. So far, we considered the noninteracting limit. Interactions can generally induce an energy mismatch between two particles localized on a  rung (provided they have opposite spin and can actually feel the onsite interaction) compared to two particles in different rungs, due to a splitting of levels that should be order of $4t_\perp^2/U$ and the emergence of new bound states. Thus, in that case, selftrapping of singlons is expected to occur. In the Appendix, we compute the entire two-body spectrum for large systems and discuss the emergent bound states and continua in the  symmetric and antisymmetric sectors. We note that this physics is   similar to the behavior of the few-magnon magnetization dynamics in  spin-1/2 ladders, see, e.g., a recent study of Heisenbeg ladders \cite{Krimphoff2017}. 

In order to illustrate that these arguments are correct, we show time- and position-dependent plots of the singlon density $\langle s_{r,\ell}\rangle$ in Fig.~\ref{fig10} for $U/t_\parallel=40$. In Fig.~\ref{fig10}(a), we present results for just two particles ($N=2$, each one placed on a site of the same rung, with opposite spin $\sigma$) in an otherwise empty lattice with $t_\perp/t_\parallel =2$. As is evident from the figure, the particles propagate with two distinct light-cones, each corresponding to two velocities, fast one being $2t_\parallel$, the second one being much smaller. The latter can be calculated from the two-particle spectrum. While the detailed derivation is  given in the Appendix, here we just present the final result, i.e., the {\it light-cone} of  the slow objects is defined by the velocity $v/t_{\parallel}=0.23$, as indicated  in Fig.~\ref{fig10}, which can be extracted from the dispersion of the respective two-body bound state. Also, we further verified that the selftrapping disappears if either $U=0$ or if the two particles have the same spin, in line with the previous arguments.

Figure~\ref{fig10}(b) illustrates the corresponding case of two singlons in the sea of doublons [compiled from the data of Fig.~\ref{fig8}(a)]. Again, as is clearly visible, the qualitative behaviour remains the same even for $t_\perp = t_\parallel$. 

\subsubsection{Holon defects}
\label{sec:holons}

\begin{figure}[!t]
\includegraphics[width=1.0\columnwidth]{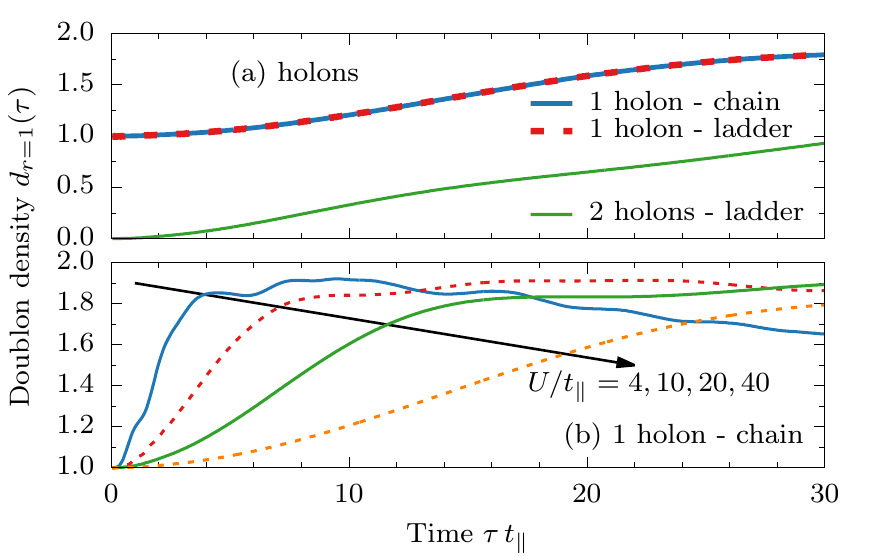}
\caption{(Color online) (a) Time dependence of the double occupancy $d_{r=1}$ on the first rung for initial states with a holon placed on the first rung, $r=1$. See the text for details on the initial state ($t_{\perp}/t_{\parallel}=0\,,1$, $L_{\mathrm{conf}}=10$, and $U/t_{\parallel}=40$). (b) Interaction-strength dependence of the single-holon escape process for a single chain (i.e., $t_{\perp}/t_{\parallel}=0$ with $L_{\mathrm{conf}}=10$).}
\label{fig11}
\end{figure}

\begin{figure*}[!t]
\includegraphics[width=1.0\textwidth]{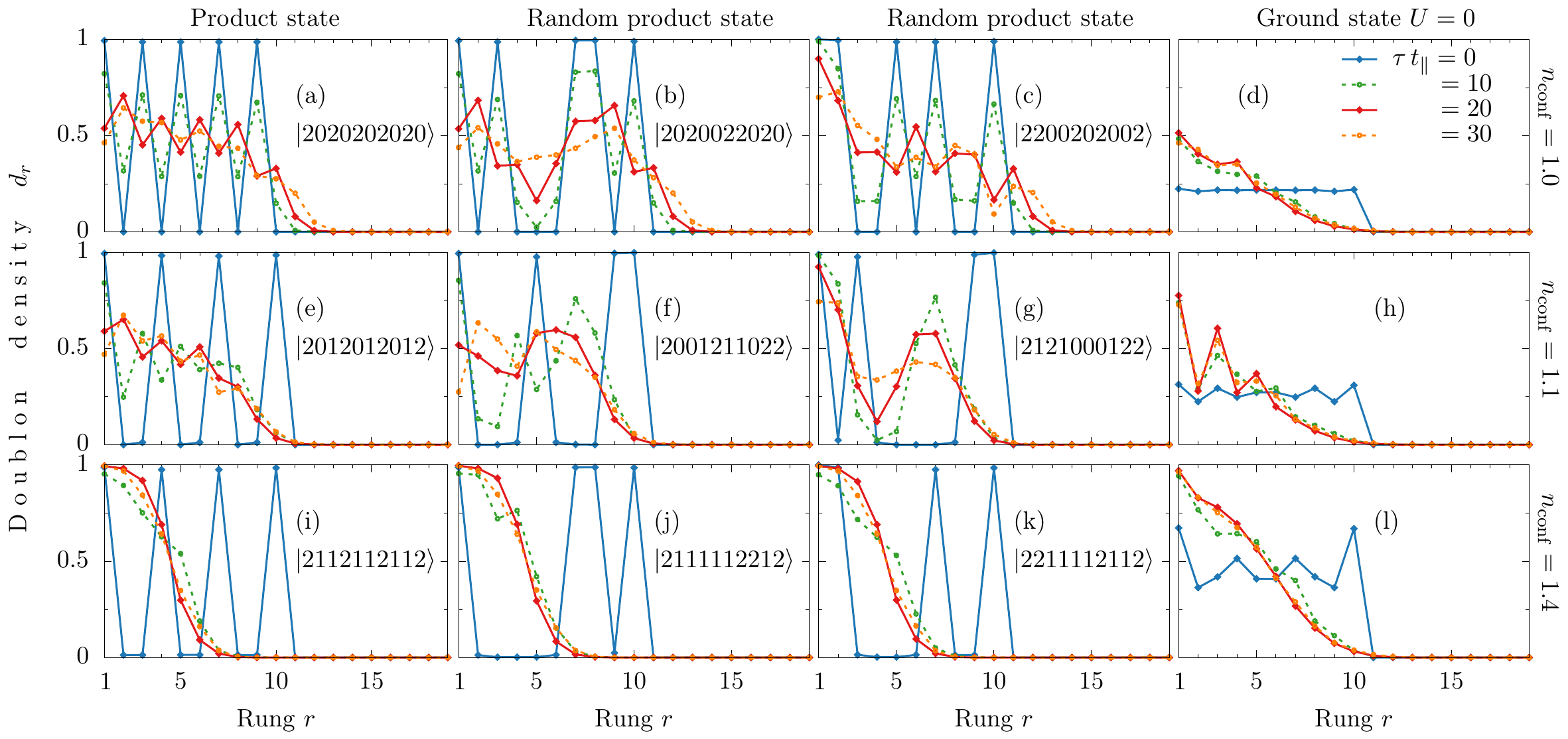}
\caption{(Color online) Snapshots of the doublon density $d_{r}$ at different times ($\tau\,t_{\parallel}=0,10,20,30$; see the legend in (d)) for various initial states as calculated for a chain with $U/t_{\parallel}=40$. The legend for all panels is depicted in panel (d). We present data for: initial states with (a-d) $n_{\mathrm{conf}}=1.0$, (e-h) $n_{\mathrm{conf}}=1.1$, and (i-l) $n_{\mathrm{conf}}=1.4$. The first column [panels (a), (e), and (i)] depicts results for product states with translationally invariant patterns, the second and third columns [panels (b-c), (f-g), and (j-k)] show data for product states with random configurations. The last column [panels (d), (h), and (l)] depicts result obtained for the initial state that is the ground state of the confined gas with $U_{\mathrm{GS}}/t_\parallel=0$ (as in Sec.~\ref{sec:distill}). }
\label{fig12}
\end{figure*}

In the last part of this section, let us comment on the dynamics of holons (i.e., empty sites) placed on the first rung. Our results presented in Fig.~\ref{fig11}(a) indicate that such a vacancy escapes the core about $\sim10$ times slower than singlons for this set of parameters [compare with Fig.~\ref{fig8}(a)]. The difference in time scales of the single-defect dynamics stems from the fact that singlons {\it exchange} their position with doublons, while in the case of a holon vacancy, the doublons have to {\it melt}, i.e., they either {dissolve} into singlons or propagate slowly. Furthermore, in contrast to singlons, the speed of escaping holons does not depend on whether there is initially one or two holons in the rung. In summary, it seems that  the presence of holons in the initial product state is the main limiting factor for an efficient quantum distillation. In the next section Sec.~\ref{sec:prod}, we will see that the effect of holons is not as drastic as the single-holon case may suggest, so long as a sufficient density of singlons is present.

Finally, in Fig.~\ref{fig11}(b), we present the interaction dependence of the single-holon dynamics on a chain. As is clearly visible, the density on the first rung increases faster for smaller interactions since there, the doublons can melt. This behaviour is consistent with the results presented in Fig.~\ref{fig6}(b). Also, our results for the time evolution of $d_{r=1}$ in the single-holon dynamics are $U$-independent on the renormalized time scale $\tau\,t^2_{\parallel}/U$ (not shown).

\subsection{Expansion from product states with average densities $n<2$}
\label{sec:prod}

In this section we investigate the efficiency of quantum distillation starting from various product states and we here restrict the analysis to the case of a chain. In the results presented below we examine two types of product states: (i) product states of holons, singlons and doublons with translational invariance in the confined region, and (ii) {\it random} product states. It is worth noting that in many experiments with ultra-cold quantum gases in optical lattices, one effectively averages over many (possibly non-identical) realizations of the initial state, while in quantum gas microscopes, the dynamics of individual chains is accessible. Here, we do not perform such an average. We rather focus on the fact that local configurations of vacancies can significantly influence the effectiveness of the process. Note that for product states, since these are not eigenstates of the Hamiltonian in the presence of the trap to begin with, there will be additional local dynamics that can result in the formation of additional doublons if two singlons of opposite spin sit on neighboring sites \cite{Vidmar2013,Bauer2015} (see also similar effects for bosons discussed in \cite{Ronzheimer2013,Vidmar2013}). 

In Fig.~\ref{fig12}, we present snapshots of the doublon density for various product states and at different times. In the same figure we present results obtained for the expansion from the $U_{\mathrm{GS}}/t_{\parallel}=0$ ground state at the same $n_{\rm conf}$ as for the product states. Note that in the former case, there is a superposition of singlons and doublons on every site. Also, during the distillation at $U/t=40$  one expects that the total doublon density remains constant, since the conservation of double occupancy is a property of the energy conserving time evolution of the Hamiltonian at large $U$, and not of the initial state. Our results presented in Fig.~\ref{fig5}(a) confirm this picture.

Let us first concentrate on product states of only doublons and holons. Figures~\ref{fig12}(a)-(c) depict results obtained from states with an equal number of doublons and holons. Consistent with our findings of Sec.~\ref{sec:imp}, we see that such initial states do not lead to any quantum distillation at all, simply because singlons are absent. Instead, doublons just slowly propagate. On the other hand, for the expansion from the ground state at $U_{\mathrm{GS}}/t_{\parallel}=0$ with the same initial density $n_{\mathrm{conf}}$, we observe an increased  doublon density in the first rungs [see Fig.~\ref{fig12}(d)]. 

Next, in Figs.~\ref{fig12}(e-g) we present results obtained from the initial state with a comparable number of doublons, singlons, and holons. Again, the presence of holons is the main limiting factor for quantum distillation. However, random product states with local clusters of singlons and doublons can lead to a large doublon density in those regions. Such a scenario is depicted in Fig.~\ref{fig12}(g), where we observe that the transient doublon density in parts of the core exceeds the value obtained for the expansion from the $U_{\mathrm{GS}}/t_{\parallel}=0$ ground state.

The data for initial product states with mixtures of holons, singlons and doublons further suggests that holons do not necessarily lead to a bottleneck for quantum distillation as one may have guessed from the discussion in Sec.~\ref{sec:holons}. Imagine a cluster in the initial state such as $|\dots 1210220\dots\rangle$. This will, within a few inverse hopping times and with a finite probability, evolve into:
\begin{eqnarray}
|\dots 1210220\dots\rangle &\rightarrow& |\dots 2101220\dots\rangle \rightarrow |\dots 2012120\dots \rangle \nonumber 
\\     & \rightarrow& |\dots 2021210\dots  \rangle \rightarrow |\dots 2022101\dots \rangle \, \nonumber
\end{eqnarray}
Therefore, the time scale for singlons to escape from the cluster is essentially unaffected by the presence of holons and all singlons can still escape, since they can exchange their positions with both singlons and doublons via first-order processes. The main effect of holons is that they can't escape themselves, thus leading to a reduction in the achievable core density after times $\propto L_{\rm conf}/(2 t_\parallel)$.

Finally, panels (i-k) depict expansions from product states with various configurations of solely singlons and doublons. Our results indicate that, after a configuration-dependent transient time $\tau\,t_{\parallel}\lesssim10$, the ensuing dynamics is practically indistinguishable. Moreover, we find that the distillation process for these product states is more efficient in comparison to the expansion from the $U_{\mathrm{GS}}/t_{\parallel}=0$ ground state, due to the (engineered) absence of any holons in the former case.

\section{Conclusions}
\label{sec:conclusions}

We studied the quantum distillation process within the Fermi-Hubbard model on a quasi-1D ladder geometry. As one of the main results of our work, we showed that this phenomenon is not limited to chains, which were previously studied \cite{Heidrich-Meisner2009}. Our investigation suggests that the distillation process on the ladder exhibits a similar dependence on model parameters such as interaction strength and initial density as in the strictly 1D case. Interestingly, for both lattices studied here, a large initial density of the core is not a necessary condition for the strong version of quantum distillation to take place. Our results indicate that even with a small initial density $n_\mathrm{conf}\sim1$, distillation occurs (provided that the initial state is constructed from a mixture of singlons and doublons). A small initial density is also preferable since it leads to the largest relative increase in the core density, making it easier to measure. The formation of a perfect band insulator, however, requires that $n\lesssim 2$ in the initial state \cite{Heidrich-Meisner2009}.

Although the overall behaviour in chains and ladders is similar, there are also some differences in the dynamics of these two setups. The most essential one is related to the time scale for the doublon density to reach its maximum inside the core. Our results indicate that due to a peculiar {\it zig-zag}--like motion of singlons, their escape dynamics on the ladder geometry is slower. A second process that gives rise to a slower escape dynamics is a partial selftrapping of singlons in the core region. These two effects are $t_{\perp}$-dependent and partially originate from the large $t_\perp>t_\parallel$ limit. Concerning the application of quantum distillation as a mechanism to produce low-entropy regions, our results are encouraging since on the ladder, the efficiency is only
slightly worse than on chains.

In order to disentangle the dynamics of different defects on top of a high-density initial state, we complemented the analysis of correlated initial states by studying the evaporation of a few initially fully localized holon or singlon defects in an otherwise band-insulating background. This result corroborates the aforementioned observations: each type of defects, i.e., singlons or holons, plays a different role in the processes of the distillation. The dynamics of singlons explains the differences in the time scales for the distillation process comparing ladders to chains. Holons (in a sea of doublons) can only move once a doublon has either partly dissolved into singlons or moved as a whole, both of which happens on much slower time scales. In a mixture of singlons, doublons and holons, the main effect of holons is primarily to reduce the achievable core density [i.e., the density reached after times $\propto L_{\rm conf}/(2t_\parallel)$], without introducing an actual bottleneck for the escape dynamics of singlons. Nevertheless, the density of holons is the main limiting factor for the quantum distillation.

In order to account for the fact that product states are frequently used in experiments, we investigated the distillation process starting from various product states. We identify configurations that lead to a fast and efficient quantum distillation, such as states with a small or vanishing number of holons. Engineered product states of doublons and singlons that can be produced with a high fidelity and thus a low admixture of holons may be the best path for observing a significant quantum distillation process, combined with a low initial average density. Employing single-site resolution techniques may be helpful in order to resolve local transient increases in the density and doublon density.

\acknowledgments
We thank T. Kohlert, H. L\"{u}schen, S. Scherg, and U. Schneider for very useful and inspiring discussions and we are indebted to M. Rigol, U. Schneider, and L. Vidmar for their comments on a previous version of the manuscript.  J.H. acknowledges support from the U.S. Department of Energy, Office of Basic Energy Sciences, Materials Science and Engineering Division. E.D. acknowledges support from the National Science Foundation, under Grant No. DMR-1404375. A.E.F. acknowledges the U.S. Department of Energy, Office of Basic Energy Sciences, for support under grant DE-SC0014407. We thank the Crete Center for Quantum Complexity and Nanotechnology for CPU time at the Metropolis cluster.

\section*{Appendix: Two-particle bound states}
\label{sec:appendix}

The formalism to calculate the two-particle spectrum in one-dimensional systems is well established \cite{Scott1994} and has been used in several scenarios involving spinless fermions, bosons, and the Hubbard chain \cite{Valiente2008,Valiente2009,Nguenang2009,Qin2008,Boschi2014,Rausch2016,Rausch2017}. We hereby extend these ideas to the case of a two-leg ladder with periodic boundary conditions in the leg direction. The approach that we present below is mostly inspired by the original work of Ref.~\onlinecite{Scott1994}, which was extended to spinfull fermions in Ref.~\onlinecite{Nguenang2009}. We start by defining a complete orthonormal basis spanned in terms of states $c^\dagger_{r,\ell,\uparrow} c^\dagger_{r^\prime,\ell^\prime,\downarrow}|0\rangle = |r,\ell;r^\prime,\ell^\prime\rangle$, where $r,r^\prime=1,\cdots,L$ are the positions of the fermions along the legs and $\ell,\ell^\prime=1,2$ are the leg indices. We exploit translational symmetry to solve the problem in subspaces with well defined lattice momentum $k=2\pi n/L$, where $n=-L/2,\cdots,L/2-1$. To this aim, we introduce the translational invariant states:
\begin{equation*}
|k,r,\ell,\ell^\prime\rangle = \frac{1}{\sqrt{L}}\sum_{d=1}^{L} e^{\imath kd}
\,\hat{T}_d\,c^\dagger_{1,\ell,\uparrow} c^\dagger_{r,\ell^\prime,\downarrow}|0\rangle\,,
\end{equation*}
where the translation operator acts on the original basis states as $\hat{T}_d|r,\ell;r^\prime,\ell^\prime\rangle = |r+d,\ell;r^\prime+d,\ell^\prime\rangle$. Periodic boundary conditions apply, implying that the indices should be assumed $\mod(L)$. In addition, we consider reflection symmetry about the plane that bisects the rungs, which leads to a new representation:
\begin{equation*}
|k,\sigma=\pm,r,\ell\rangle = \frac{1}{\sqrt{2}}\Big(|k,r,1,\ell\rangle \pm |k,r,2,3-\ell\rangle \Big)\,.
\end{equation*}
The new states are classified by lattice momentum $k$, symmetric/antisymmetric (bonding/antibonding) $\sigma=\pm$ sectors, and relative position between the two fermions. Note that $\sigma=\pm$ can be associated to lattice momenta $0,\pi$.

The Hamiltonian matrix elements are easy to obtain as:
\begin{eqnarray*}
H |k,\sigma,\ell,r\rangle =
&-& t_\parallel \Big(q|k,\sigma,\ell+1,r\rangle + q^*|k,\sigma,\ell-1,r\rangle \Big)\\
&-& 2t_\perp \delta_{\sigma,+}|k,\sigma,\ell,3-r\rangle +
U\delta_{(\ell,r),0}|k,\sigma,0,0\rangle\,,
\end{eqnarray*}
with $q=\cos{(k/2)}\exp{(\imath k)}$ and $|k,1,\sigma,r\rangle \equiv |k,L+1,\sigma,r\rangle$. Notice that the rung hopping has no effect on the antibonding states, which implies the lack of bound states in this sector. The use of symmetries enables us to numerically solve the spectrum for very large systems (hundreds of sites) since the matrix dimension for each subspace $(k,\sigma)$ is $dim=2L$.

\begin{figure}[!t]
\includegraphics[width=1.0\columnwidth]{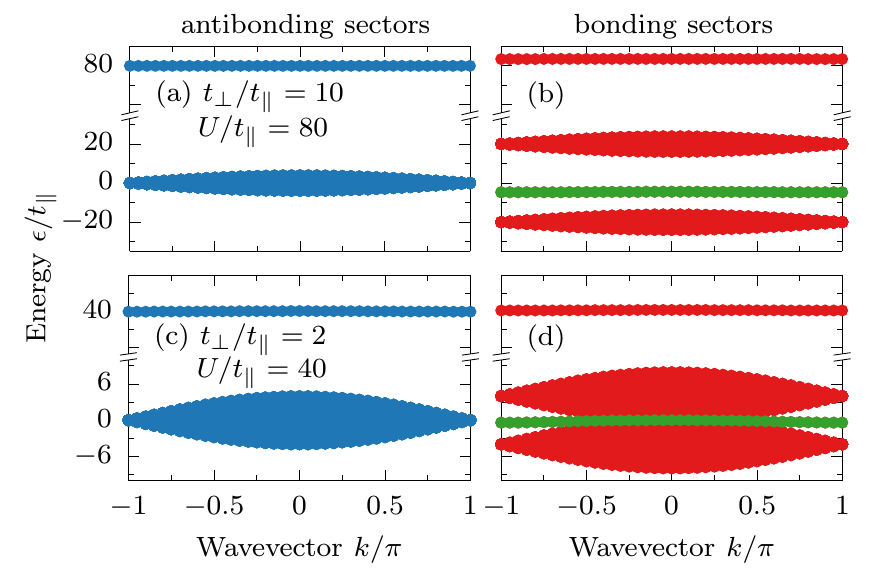}
\caption{Two-particle states in the (a),(c) antibonding/antisymmetric  and (b), (d) bonding/symmetric   sector of the spectrum for (a,b) $t_\perp/t_\parallel=10\,,U/t_\parallel=80$ and (c,d) $t_\perp/t_\parallel=2\,,U/t_\parallel=40$ as calculated for $L=40$. Green points in the right panel depict bound states.}
\label{figapp}
\end{figure} 

In order to intuitively understand the two-particle spectrum we first analyze the case $t_\perp/t_\parallel=10$ and $U/t_\parallel=80$, which brings us close to the isolated rung limit. We show the spectrum for such  parameters in Fig.\ref{figapp} (a,b) for a relatively small ladder with $L=40$ for visualization purposes. In the antisymmetric/antibonding sector we encounter a flat band at $\omega/t_{\parallel} = U/t_{\parallel}$ and a scattering continuum centered about $\omega/t_{\parallel} \sim 0$. One can show that in this sector there is no hybridization between states with single and double occupation, and the latter are localized and non-dispersive. In the symmetric/bonding sector we distinguish bound states centered approximately around $\omega \sim -4t_\perp^2/U$ and $U+4t_\perp^2/U$. These states can hop coherently along the leg direction. In addition, we find two continua of scattering states around $\omega \sim \pm 2t_\perp$, which correspond to two independent fermions on separate rungs far apart from each other.

Results for the parameters close to the ones used in this work (i.e., $t_\perp/t_\parallel=2,U/t_\parallel=40$) are shown in panels (c) and (d) of Fig.~\ref{figapp}. The fermions are now more dispersive, increasing the bandwidth of the scattering continua. One can clearly resolve the band corresponding to the low-energy bound states, which can be accurately fitted to an expression of the form
\begin{equation*}
\epsilon(k)=\alpha \cos{(k)} - \beta \cos{(2k)}\,.
\end{equation*}
The maximum slope of this dispersion yields the value $v/t_\parallel \sim 0.23$.

\bibliography{references}
\end{document}